\documentclass[twocolumn,showpacs,preprintnumbers,amsmath,amssymb]{revtex4}

\usepackage{graphicx}
\usepackage{subfigure}
\usepackage{dcolumn}
\usepackage{bm}

\def\mt2{\ensuremath{m_{T2}^H}}

\begin{document}


\title{Transverse mass observables for charged Higgs searches at hadron colliders}

\author{Eilam Gross}
 \email{eilam.gross@weizmann.ac.il}
\author{Ofer Vitells}%
 \email{ofer.vitells@weizmann.ac.il}
\affiliation{%
Weizmann Institute of Science\\
Rehovot, Israel
}%

\date{\today}

\begin{abstract}
Charged Higgs boson may be produced at hadron colliders via the
decay of top quarks. At the LHC, The large production cross section
of top quarks pairs will make it one of the earliest channels
allowing to search for physics beyond the standard model. Assuming
the decay $H^+ \rightarrow \tau^+\nu$, previous searches have so far
considered only hadronic $\tau$ decays. Here we show that by using
appropriate kinematical variables (transverse mass observables),
leptonic $\tau$ decays can be used as well, for both observing and
possibly measuring the mass of the charged Higgs boson. This can
increase the overall experimental sensitivity to charged Higgs
bosons at hadron colliders.
\end{abstract}

\maketitle

\section{\label{sec:intro}Introduction}
Charged Higgs bosons arise in models with extended Higgs sector such
as two Higgs doublets models (2HDM) and in particular the Minimal
Supersymmetric Standard Model (MSSM). If the charged Higgs mass is
below the top mass, It is expected to be produced at hadron
colliders dominantly via top quark decay, due to the large Yukawa
coupling of the top. At the LHC, the large production cross section
of top quark pairs means that the existence of charged Higgs bosons
could be probed with early data, with a potential of being among the
first evidences of physics beyond the standard model. Here, we
present methods for observing such a charged Higgs and possibly
measuring it's mass, assuming that it decays to a $\tau$ lepton and
a neutrino, $H^+ \rightarrow \tau^+\nu$, and using the leptonic
decay modes of the $\tau$, either $\tau^+ \rightarrow
e^+\nu_e\bar{\nu_{\tau}}$ or $\tau^+ \rightarrow
\mu^+\nu_\mu\bar{\nu_{\tau}}$. The decay mode $H^+ \rightarrow
\tau^+\nu$ dominates in the MSSM for $\tan\beta > 1$ and is
therefore a favorable search channel. Previous searches at the
Tevatron \cite{cdf}\cite{d0} considered only hadronic $\tau$ decays,
since those allow the $\tau$ lepton to be identified and an excess
of $\tau$ leptons over the standard model prediction can provide a
direct evidence for a charged Higgs boson. On the other hand, the
leptonic decay modes involving an isolated electron or muon have
experimental advantages with regard to triggering the event and
suppressing backgrounds, and they are expected to be associated with
less systematic uncertainties compared to the hadronic mode. This
makes these channels particularly important for early stages of the
LHC.

The event topology we consider here belongs to a wide class of
events where the presence of undetectable particle(s) in the final
states prevents full reconstruction of the sought after particle's
mass. It is often possible in such cases to define kinematical
variables that are bounded by the unknown mass, and therefore it's
value can be inferred from an edge in the distribution of that
variable over many events. We generally refer to such variables as
transverse mass observables, following the most known example which
is the $W$ transverse mass in the process $W \rightarrow l\nu$. Many
generalizations to more complicated processes have been studied in
past years, such as the $m_{T2}$ variable \cite{mt2} and others
\cite{others1}\cite{others2}\cite{others3}. In the following
sections we apply this approach to charged Higgs searches in
$t\bar{t}$ events. We consider both semi-leptonic and di-leptonic
final states, and show that such observables could be constructed in
those cases and provide valuable information for observing and
measuring the mass of the charged Higgs boson.

\subsection*{Event simulation}

To study the performance of the proposed variables we use MADGRAPH
\cite{mad} for event generation, interfaced to PYTHIA \cite{pythia}
for showering and hadronization. $\tau$ decays are handled by TAUOLA
\cite{tauola}. Detector effects and jet clustering are done with the
PGS \cite{pgs} fast simulation package , configured with LHC
parameters. All samples are generated for a center of mass energy of
14 TeV.

\section{Semi-leptonic $t\bar{t}$ events}

\begin{figure}[h!]
\begin{center}
\includegraphics*[width=8cm]{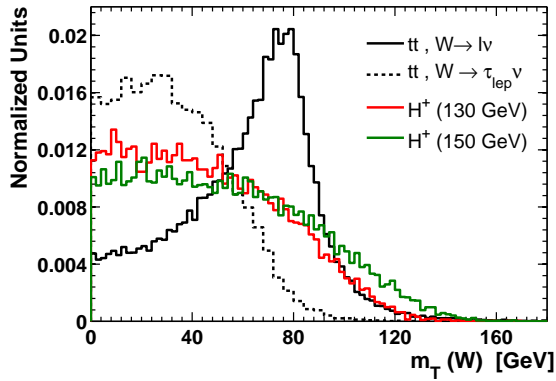}
\end{center}
\caption{The W transverse mass distributions for leptonic W decays
and charged Higgs boson decays with $m_{H^ +  } = 130,150$ GeV. all
distributions are normalized to unit area. A minimum cut of 20 GeV
for the lepton transverse momentum is applied. } \label{fig:mtw}
\end{figure}

In this section we consider events in which one of the top quarks
decays to a charged Higgs  ($t \rightarrow bH$) while the second top
quark decays hadronically and is therefore fully reconstructable ($t
\rightarrow bW \rightarrow bjj$). This is the most useful channel
for top reconstruction since the missing energy and isolated lepton
together with b-tagged jets provide a good rejection against
backgrounds. The main background to this process is the standard
model decay $t \rightarrow Wb$ with a subsequent leptonic decay of
the W, either directly $W \rightarrow \ell\bar{\nu}$ or via a $\tau$
: $W \rightarrow \tau\bar{\nu} \rightarrow
\ell\bar{\nu}\nu\bar{\nu}$. It is natural for this channel to
reconstruct the W transverse mass, defined as \cite{bib:wmt}
\begin{equation}
(m_T^W )^2  = 2 p_T^\ell  p_T^{miss} (1 - \cos \phi_{\ell,miss} )
\label{eq:mtw}
\end{equation}
\noindent where $p_T^\ell$,$p_T^{miss}$ are the lepton and missing
transverse momenta respectively and $\phi_{\ell,miss}$ is the
azimuthal angle between them. For direct W decays this transverse
mass has a Jacobian peak near the W mass, hence it can provide
separation between the two $W$ decay modes, as well as charged Higgs
decays (see Figure \ref{fig:mtw}).

In principle, the presence of charged Higgs boson decays will be
manifested as an excess of events in the lower region of the
transverse mass distribution. In practice however, to observe such
an excess would require an accurate modeling of the shape of the
distribution which depends on the missing momentum resolution. This
may be hard to achieve. A second drawback of this approach is that
almost no information about the \emph{mass} of the charged Higgs
boson can be obtained. Still, by requiring $m_T^W$ to be
sufficiently below the peak region, one can reject a large part of
the background and enhance the signal content of the sample.

It is useful to define the W transverse mass (\ref{eq:mtw})
equivalently as a minimization result, i.e.
\begin{equation}
(m_T^W )^2  = \mathop {\min}\limits_{\left\{ {\scriptstyle
p_z^{miss} ,E^{miss}  \hfill \atop
  \scriptstyle (p^{miss})^2  = 0  \hfill} \right\}} {\kern 1pt} [(p^\ell   + p^{miss} )^2 ]
\label{eq:mtw1}
\end{equation}
\noindent where $p^\ell,p^{miss}$ are the lepton and neutrino
(missing) four momenta, respectively. It is straightforward to show
that the above requirement indeed leads to the expression
(\ref{eq:mtw}) (see a similar derivation in appendix A). The
definition (\ref{eq:mtw1}) makes explicit the fact that $m_T^W \leq
m_W$, i.e. $m_T^W$ is bounded by the $W$ mass. Furthermore, it
implies that $m_T^W$ is the \emph{best} bound that can be placed on
$m_W$, per event.

For the case of a leptonic $\tau$ decay (either from a W or charged
Higgs boson) the constraint used in (\ref{eq:mtw1}) is not valid
since the missing momentum comes from three neutrinos and therefore
$(p^{miss})^2 \neq 0$. However, if one of the two b-jets in the
event could be associated with the leptonically decaying top quark
then the on-shell constraint for the top quark could be used, and
one would define analogously a ``Charged Higgs transverse mass'' in
the following way:

\begin{equation} (m_T^H )^2  = \mathop {\max}\limits_{\left\{
{\scriptstyle p_z^{miss} ,E^{miss}  \hfill \atop
  \scriptstyle (p^{miss} + p^\ell + p^b)^2  = m_{top}^2  \hfill} \right\}} {\kern 1pt} [(p^\ell   + p^{miss} )^2 ]
\label{eq:mth1}
\end{equation}

\noindent note that in this case the transverse mass is defined by
maximization of the invariant mass, since it is bounded from above
by the top quark mass. The charged Higgs transverse mass by
definition therefore satisfies $m_{H^ +  }  \le m_T^H  \le m_{top}$,
where $m_{H^ +  }$ is the true charged Higgs mass and $m_{top}$ is
the nominal value of the top mass used in the constraint. The
explicit expression for it can be easily derived and is given by:
(see appendix A)

\begin{equation}
(m_T^H )^2  = \left(\sqrt {m_{top}^2  + (\vec p_T^\ell   + \vec
p_T^b + \vec p_T^{miss} )^2 }  - p_T^b \right)^2  - \left(\vec
p_T^\ell + \vec p_T^{miss}\right)^2 \label{eq:mth2}
\end{equation}

It should be noted that top quark decay can involve gluon radiation
\cite{bib3}. In such case the constraint used in (\ref{eq:mth1}) is
not exact. If the top quark emits a gluon and turns off-shell before
it's decay, then the invariant mass of the decay products is smaller
then the top pole mass
\begin{equation}
\sqrt {(p^\ell   + p^b  + p^{miss} )^2 }  = m_{top}^ *   \le m_{top}
\end{equation}

\noindent however, the charged Higgs transverse mass (\ref{eq:mth2})
is a monotonically increasing function of the input value of
$m_{top} $ and therefore the following relation still holds:
\begin{equation}
m_T^H (m_{top} ) \ge m_T^H (m_{top}^ *  ) \ge m_{H^ +  }^{}
\end{equation}

\noindent i.e., $m_T^H $ retains the property of being bounded by
the charged Higgs mass also in the presence of gluon radiation. The
values of $m_{H^+}$ however will be shifted upwards, making the
Jacobian peak less pronounced.

\subsection*{Event reconstruction}

To reconstruct the charged Higgs transverse mass, one must correctly
associate a b-jet to the leptonic top decay. For the purpose of our
study we consider events where exactly two of the jets are b-tagged,
such that there are two possible choices. The selection cuts we
apply are the following:
\begin{itemize}
\item At least four jets with $p_T > 20$ GeV and $|\eta|<2.5$
\item Two of the above jets are b-tagged
\item Exactly one isolated lepton ($e$ or $\mu$) with $p_T > 20$
\end{itemize}
To choose between the two possible b-jet assignments, we first
consider the invariant mass $m_{\ell b}^2 = (p^{\ell}+p^b)^2$. For
the correct b-jet, this is kinematically bounded by
\begin{equation}
m_{\ell b}^2 \leq m_{top}^2 - m_W^2 \label{eq:mlb}
\end{equation}
\noindent note that this bound holds for both $W$ and charged Higgs
boson decays, since we are assuming $m_{H^+} \geq m_W$ \cite{lep}.
In case that one of the lepton-b pairings violates the above
condition, the opposite pairing in selected. In events where both
pairings satisfy (\ref{eq:mlb}), we further consider the top quark
transverse mass :
\begin{equation}
(m_T^{top})^2 = m_{\ell b}^2 + 2(E_T^{\ell b} p_T^{miss} - \vec
p_T^{\ell b}\cdot \vec p_T^{miss}) \label{eq:mTtop}
\end{equation}
which is required to satisfy $m_T^{top} \leq m_{top}$. Again, if one
of the pairings violates this condition then the opposite one is
selected. Finally, if this is not the case, the hadronic top is
reconstructed using two of the light (non b-tagged) jets plus one of
the b-jets. The b-jet that results in the invariant mass $m_{jjb}$
being closest to the nominal value of $m_{top}$ is associated to the
hadronic top, and thus the other b-jet is associated to the leptonic
top.

\begin{figure}[h!]
\begin{center}
\includegraphics*[width=8cm]{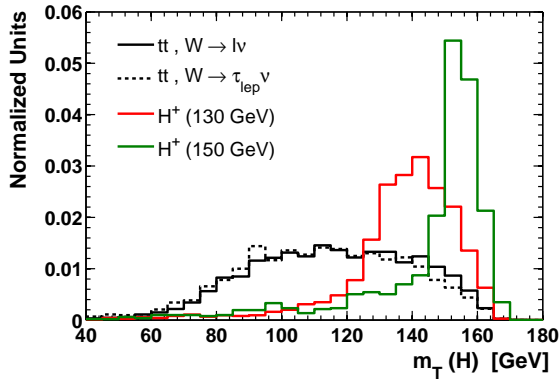}
\end{center}
\caption{The charged Higgs transverse mass distributions for
leotonic W decays and a charged Higgs decays with $m_{H^ +  } =
130,150$ GeV. all distributions are normalized to unit area.}
\label{fig:mth}
\end{figure}

Figure (\ref{fig:mth}) shows the charged Higgs transverse mass
distributions after selection and reconstruction. The power of
$m_T^H$ as a discriminant based on the mass difference is clearly
observed.

A useful property of $m_T^H$ is that the background distribution is
independent of the $W$ decay mode (direct or $\tau$-mediated). This
is easily understood from (\ref{eq:mth2}), which depends on the
lepton and missing transverse momenta only via their sum $\vec
p_T^\ell + \vec p_T^{miss} = \vec p_T^{W}$.

\begin{figure}[h!]
\begin{center}
\includegraphics*[width=8cm]{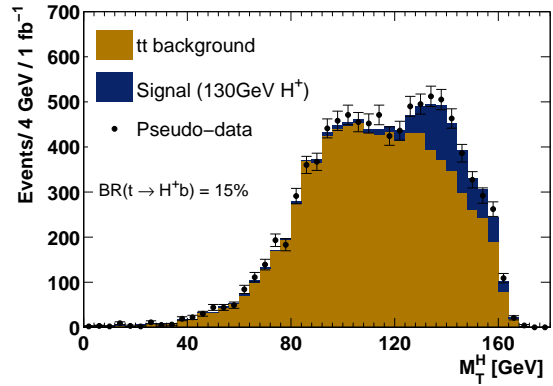}
\end{center}
\caption{Example pseudo-experiments (points with error bars) on top
of the signal and background distributions, assuming $Br(t
\rightarrow H^+b)=15\%$ and $m_{H^+}=130$ GeV, for an integrated
luminosity of 1 fb$^{-1}$.  } \label{fig:exp}
\end{figure}

In figure \ref{fig:exp} an example pseudo-experiment containing both
the main background and a charged Higgs signal with $m_{H^+}=130$
GeV is shown. The number of events correspond to an integrated
luminosity of 1 fb$^{-1}$ at $\sqrt{s}=14$ TeV, and the branching
ratio $Br(t \rightarrow H^+b)$ is assumed to be 15\%, roughly the
current upper limit set by the Tevatron searches \cite{d0}. The
statistical significance of discovery in this case exceeds
$5\sigma$, demonstrating the potential power of this variable.

\section{dileptonic $t\bar{t}$ events}

\begin{figure}[h!]
\begin{center}
\includegraphics*[width=7cm]{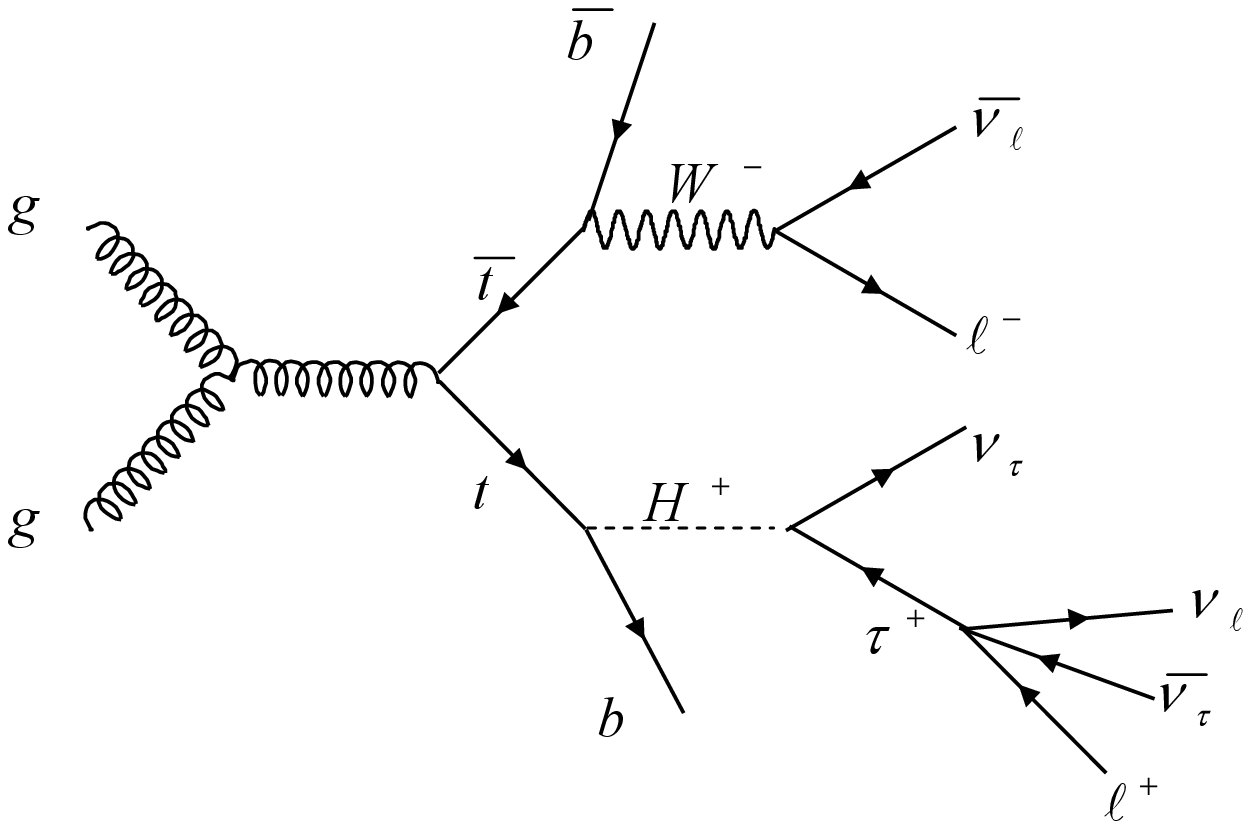}
\end{center}
\caption{A Feynman diagram illustrating the event topology of a
di-leptonic $t\bar{t}$ event involving a charged Higgs decay. }
\label{fig:feyn_dilep}
\end{figure}

In this section we generalize the above procedure for dileptonic
$t\bar{t}$ events, as depicted in Figure \ref{fig:feyn_dilep}. In
this case the final state includes two leptons and missing energy on
both sides of the event, making the reconstruction of the event more
complicated.

To define an analogue to the charged Higgs transverse mass of the
previous section, we follow the same procedure, i.e. maximize the
invariant mass subject to all available constraints that are given
by the known $W$ and top quark masses, as well the conservation of
momentum in the transverse plane. We have the following set of six
constraints, corresponding to the notation of Figure
\ref{fig:feyn_dilep} :
\begin{eqnarray}
\label{eq:cons}
(p^{H^+} + p^b)^2 = m_{top}^2 \nonumber\\
(p^{\ell^-} + p^{\bar{\nu}_\ell})^2 = m_W^2 \nonumber\\
(p^{\ell^-} + p^{\bar{\nu}_\ell} + p^{\bar{b}})^2 = m_{top}^2 \\
(p^{\bar{\nu}_\ell})^2 = 0 \nonumber\\
\vec p_T^{\hspace{1mm}H^+} - \vec p_T^{\hspace{1mm}\ell^+} + \vec
p_T^{\hspace{1mm}\bar{\nu}_\ell} = \vec p_T^{\hspace{1mm}miss}
\nonumber
\end{eqnarray}

Here $p^{H^+}$ and $p^{\bar{\nu}_\ell}$ represent the unknown
quantities of the event. Note that we have not specified constraints
for the particles that are the decay products of the charged Higgs,
since this would not supply us with additional information that can
be use to constrain the charged Higgs mass. The system of
constraints (\ref{eq:cons}) therefore amounts to two free parameters
(compared to only one in the semi-leptonic case) over which we
maximize the charged Higgs mass, to obtain the variable $m_{T2}^H$
\begin{equation}
(m_{T2}^H)^2 = \mathop {\max }\limits_{\left\{ \scriptstyle
constraints \right\}}[(p^{H^+})^2 ]
\end{equation}

If we choose one of the free parameters to be the $z$-component of
the charged Higgs momentum, we can immediately perform the
maximization over it using the result of the previous section, and
so $m_{T2}^H$ is equivalent to
\begin{equation}
m_{T2}^H = \mathop {\max }\limits_{\left\{ \scriptstyle constraints
\right\} } {\kern 1pt} [m_T^H(\vec p_T^{H^+})] \label{eq:mt2h}
\end{equation}
with
\begin{equation}
(m_T^H(\vec p_T^{H^+}) )^2  = \left(\sqrt {m_{top}^2  + (\vec
p_T^{H^+} + \vec p_T^b)^2 }  - p_T^b \right)^2  - \left(\vec
p_T^{H^+}\right)^2 \label{eq:mth2d}
\end{equation}

The maximization over the remaining parameter needs to be performed
numerically. A computation procedure which allows for this
maximization to be easily performed is given in appendix B.

\subsection*{Selection and reconstruction}

\begin{figure}[ht!]
\begin{center}
\includegraphics*[width=8cm]{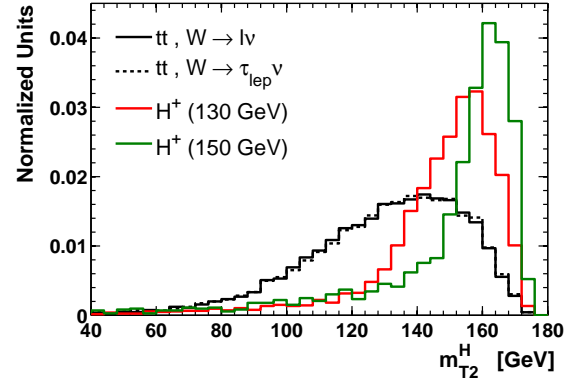}
\end{center}
\caption{distribution of $m_{T2}^H$ for dileptonic $t\bar{t}$ events
containing a charged Higgs with $m_{H^+}=130,150$ GeV (red and
green) and background events with a $W$ boson decays
(black)}\label{fig:mt2_wh}
\end{figure}

The selection cuts we apply in this case are similar to those used
for the semi-leptonic channel, except that we require an additional
isolated lepton with $p_T>15$ GeV. The two leptons and two b-jets
need to be paired according to the two corresponding top decays in
order for $m_{T2}^H$ to be calculated. To do this we calculate the
invariant mass $m_{\ell b}^2$ for each possible pair and require it
to satisfy the bound (\ref{eq:mlb}). We keep only events where this
requirement resolves the ambiguity, i.e. events where exactly one
pairing satisfies (\ref{eq:mlb}). This keeps only about 50\% of the
events, however the accuracy of the assignment is high, at about
95\%. The calculation of $m_{T2}^H$ further requires assigning one
of the two leptons to the charged Higgs decay. Here we always choose
it as the one with lower transverse momentum. The resulting
distributions of $m_{T2}^H$ are shown in Figure \ref{fig:mt2_wh}.
The effect of having additional free parameter clearly worsens the
situation compared to the semi-leptonic case, however there is still
a clear discrimination between the W and charged Higgs bosons.

\begin{figure}[h!]
\begin{center}
\includegraphics*[width=8cm]{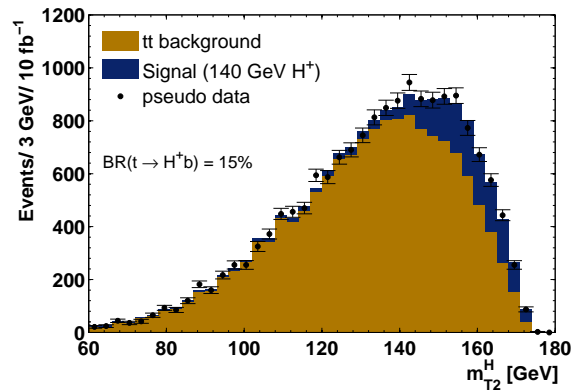}
\end{center}
\caption{Example pseudo-experiments (points with error bars) on top
of the signal and background distributions, assuming $Br(t
\rightarrow H^+b)=15\%$ and $m_{H^+}=140$ GeV, for an integrated
luminosity of 10 fb$^{-1}$.  } \label{fig:exp2}
\end{figure}

In figure \ref{fig:exp2} an example pseudo-experiment containing
both the main background and a charged Higgs signal with
$m_{H^+}=140$ GeV is shown. Here the number of events correspond to
an integrated luminosity of 10 fb$^{-1}$ at $\sqrt{s}=14$ TeV, and
the branching ratio $Br(t \rightarrow H^+b)$ is assumed to be 15\%.
In this case, since the kinematic edge is less pronounced, to
observe a charged Higgs using the $m_{T2}^H$ variable requires a
rather accurate knowledge of the background shape. The excess of
events above the charged Higgs mass would then provide an indication
for the charged Higgs in this channel, complementary to the
semi-leptonic one.

\section{Conclusions}
We have demonstrated how the leptonic decay channel of the $\tau$
could be used as a search channel for a charged Higgs boson at
hadron colliders. The charged Higgs transverse mass defined for
semi-leptonic $t\bar{t}$ events, as well as the analogous
generalization for di-leptonic events, are sensitive to the charged
Higgs mass and could therefore be used as discriminating variables
that would give evidence to such a particle, and a way to measure
it's mass. This would provide an important complementary measurement
to methods based on hadronic $\tau$ identification, and an
enhancement to the overall discovery sensitivity.

\section*{Acknowledgements}
One of us (E.G.) is obliged to the Benoziyo center for High Energy
Physics, to the the Israeli Science Foundation(ISF), the Minerva
Gesellschaft and the German Israeli Foundation (GIF) for supporting
this work.

\appendix

\section{Derivation of $m_T^H$}
In this appendix, we derive the expression for the charged Higgs
transverse mass (\ref{eq:mth2}). Our goal is to maximize the
expression of invariant mass:

\begin{equation}
 m_{H^ +  }^2  = (p^\ell   + p^{miss} )^2
\end{equation}

\noindent with respect to the two unknown components of $p^{miss} $
, while holding the constraint:
\begin{equation}
m_{top}^2  = (p^{miss}  + p^\ell   + p^b )^2
\end{equation}

\noindent For convenience, we adopt the notation $p_\parallel   =
(E,p_z )$ , satisfying $p_\parallel ^2  = E^2  - p_z^2 $ ,and $\vec
p_T  = (p_x ,p_y )$ with $\vec p_T^2  = p_x^2  + p_y^2 $ . Equation
(A.1) and (A.2) can be re-written as:

\begin{equation}
m_{H^ +  }^2  = (p_\parallel ^\ell   + p_\parallel ^{miss} )^2  -
(\vec p_T^\ell   + \vec p_T^{miss} )^2
\end{equation}

\begin{equation}
m_{top}^2  = (p_\parallel ^\ell   + p_\parallel ^{miss}  +
p_\parallel ^b )^2  - (\vec p_T^\ell   + \vec p_T^{miss}  + \vec
p_T^b )^2
\end{equation}

\noindent To maximize Eq. (A.3), we introduce a Lagrange
multiplier$\lambda $  and differentiate with respect to $p_\parallel
^{miss} $ :

\begin{equation}
\frac{\partial }{{\partial p_\parallel ^{miss} }}((p_\parallel ^\ell
+ p_\parallel ^{miss} )^2  - (\vec p_T^\ell   + \vec p_T^{miss} )^2
) = 0
\end{equation}

\noindent This gives
\begin{equation}
p_\parallel ^{miss}  = \frac{\lambda }{{1 - \lambda }}p_\parallel
^\ell   - p_\parallel ^b
\end{equation}

\noindent plugging this into Eq. (A.4) we obtain for $\lambda $ :
\begin{equation}
1 - \lambda  = \frac{{p_T^b }}{{\sqrt {m_{top}^2  + (\vec p_T^\ell +
\vec p_T^{miss}  + \vec p_T^b )^2 } }}
\end{equation}

\noindent where we have approximated the b-quark to be massless.
Plugging (A.7) and (A.6) into (A.3) we obtain the final result:
\begin{equation}
(m_T^H )^2  = (\sqrt {m_{top}^2  + (\vec p_T^\ell   + \vec p_T^b  +
\vec p_T^{miss} )^2 }  - p_T^b )^2  - (\vec p_T^\ell   + \vec
p_T^{miss} )^2
\end{equation}

\noindent It should be noted that exactly the same procedure could
be applied to the case of a leptonic W decay $W \to \ell \nu $ , by
replacing the constraint (A.2) with $(p^{miss} )^2  = 0$
 . This will result in the usual expression for the W transverse mass.

\section{Computation procedure for \mt2} \label{appx}
In this section we describe the computational procedure by which the
maximization of Eq. \ref{eq:mt2h} is performed. The transverse
component of the charged Higgs momentum $\vec p_T^{H^+}$ is
constrained by the set of equations

\begin{eqnarray}
\label{eq:cons2} \vec p_T^{\hspace{1mm}H^+} - \vec
p_T^{\hspace{1mm}\ell^+} + \vec p_T^{\hspace{1mm}\bar{\nu}_\ell} =
\vec p_T^{\hspace{1mm}miss} \nonumber\\
(p^{\bar{\nu}_\ell})^2 = 0 \nonumber\\
(p^{\ell^-} + p^{\bar{\nu}_\ell})^2 = m_W^2 \\
(p^{\ell^-} + p^{\bar{\nu}_\ell} + p^{\bar{b}})^2 = m_{top}^2
\nonumber
\end{eqnarray}

We wish to obtain a parametrization of $\vec p_T^{H^+}$ as a
function of a single variable such that the above constraints are
satisfied. To do this we rewrite the last two equations as the
following linear matrix equation:
\begin{equation}
\mathbf{A_1} \eta + \mathbf{A_2} \xi = m \label{eq:xi}
\end{equation}

where
\\

$\hfill \mathbf{A_1} = \left( \begin{smallmatrix} E^{\bar{b}} &
-p_z^{\bar{b}}\\E^{\ell^-} & -p_z^{\ell^-} \end{smallmatrix}
\right),\hfill \mathbf{A_2} = \left(\begin{smallmatrix}
-p_x^{\bar{b}} & -p_y^{\bar{b}}\\-p_x^{\ell^-} & -p_y^{\ell^-}
\end{smallmatrix} \right) \\
\hfill \eta = \left( \begin{smallmatrix} E^{\bar{\nu}_\ell} \\
p_z^{\bar{\nu}_\ell}\end{smallmatrix} \right),\hfill
\xi = \left( \begin{smallmatrix} p_x^{\bar{\nu}_\ell} \\
p_y^{\bar{\nu}_\ell}\end{smallmatrix} \right),\hfill
m = \frac{1}{2}\left( \begin{smallmatrix} m_{top}^2 - m_W^2 - m_{\ell b}^2 \\
m_W^2\end{smallmatrix} \right)$

and

\begin{equation}
m_{\ell b}^2 = (p^{\bar{b}}+p^{\ell^-})^2
\end{equation}

The zero-mass constraint of the neutrino is given in this notation
by
\begin{equation}
\eta^T g \eta - \xi^T\xi = 0
\end{equation}

where g is the 1+1 Lorentz metric, $g = \left( \begin{smallmatrix}
1&0\\0&-1 \end{smallmatrix} \right)$.

using (\ref{eq:xi}) this can be written as:

\begin{equation}
\xi^T (\mathbf{1} - \mathbf{A}^Tg\mathbf{A})\xi + 2\xi^T
\mathbf{A}^T g \tilde{m} - \tilde{m}^Tg\tilde{m} = 0
\end{equation}

where $\mathbf{A} \equiv \mathbf{A_1}^{-1}\mathbf{A_2}$ and
$\tilde{m} \equiv \mathbf{A_1}^{-1} m$.

The matrix $\mathbf{1} - \mathbf{A}^Tg\mathbf{A}$ is symmetric and
can be shown to be positive-definite. It can therefore be decomposed
in the following way:
\begin{equation}
\mathbf{1} - \mathbf{A}^Tg\mathbf{A} = \mathbf{L}^T \mathbf{L}
\end{equation}

e.g. via a Cholesky decomposition. Using this we obtain

\begin{equation}
|\mathbf{L} \xi + \omega| = \sqrt{\omega^T\omega +
\tilde{m}^Tg\tilde{m}} \equiv \sqrt{Q}
\end{equation}

where $\omega = (\mathbf{L}^T)^{-1} \mathbf{A}^T g \tilde{m}$. \\
This is solved by
\begin{equation}
\xi = \mathbf{L}^{-1}(\sqrt{Q}\hat{r}(\phi) - \omega)
\end{equation}
where $\hat{r}$ is the unit vector in the transverse plane,\\
 $\hat{r}(\phi) = \left( \begin{smallmatrix} \sin\phi \\
\cos\phi \end{smallmatrix} \right)$.\\
We can therefore express  $\vec p_T^{H^+}$ as a function of $\phi$:
\begin{equation}
\vec p_T^{\hspace{1mm}H^+} = \vec p_T^{\hspace{1mm}miss} + \vec
p_T^{\hspace{1mm}\ell^+} - \xi(\phi)
\end{equation}
such that
\begin{equation}
m_{T2}^H = \mathop {\max }\limits_{ \phi  } {\kern 1pt}
[m_T^H(\phi)] \label{eq:mt2_2}
\end{equation}
The maximization over $\phi$ can now  be performed by any standard
function minimization algorithm, or by scanning over the range $0 <
\phi < 2\pi$. The result is the generalized charged Higgs transverse
mass.\\

%
%
\newpage 

\end{document}